\documentclass[a4,12pt]{article}
\usepackage[dvips]{graphicx,psfrag}
\usepackage{float}
\usepackage{amsmath,amsthm,amssymb}
\def\ls{\lower0.5ex\hbox{$\buildrel >\over{\scriptstyle\sim}$}}
\def\rs{\lower0.5ex\hbox{$\buildrel <\over{\scriptstyle\sim}$}} 
\allowdisplaybreaks
\begin{document}
\pagestyle{empty} \setlength{\footskip}{2.0cm}
\setlength{\oddsidemargin}{0.5cm}
\setlength{\evensidemargin}{0.5cm}
\renewcommand{\thepage}{-- \arabic{page} --}
\def\mib#1{\mbox{\boldmath $#1$}}
\def\bra#1{\langle #1 |}  \def\ket#1{|#1\rangle}
\def\vev#1{\langle #1\rangle} \def\dps{\displaystyle}
\newcommand{\fcal}{{\cal F}}
\newcommand{\gcal}{{\cal G}}
\newcommand{\ocal}{{\cal O}}
\newcommand{\El}{E_\ell}
\renewcommand{\thefootnote}{$\sharp$\arabic{footnote}}
\newcommand{\W}{{\scriptstyle W}}
 \newcommand{\I}{{\scriptscriptstyle I}}
 \newcommand{\J}{{\scriptscriptstyle J}}
 \newcommand{\K}{{\scriptscriptstyle K}}
%
% ------------------------------------------------------------
 \def\thebibliography#1{\centerline{REFERENCES}
 \list{[\arabic{enumi}]}{\settowidth\labelwidth{[#1]}\leftmargin
 \labelwidth\advance\leftmargin\labelsep\usecounter{enumi}}
 \def\newblock{\hskip .11em plus .33em minus -.07em}\sloppy
 \clubpenalty4000\widowpenalty4000\sfcode`\.=1000\relax}\let
 \endthebibliography=\endlist
 \def\sec#1{\addtocounter{section}{1}\section*{\hspace*{-0.72cm}
 \normalsize\bf\arabic{section}.$\;$#1}\vspace*{-0.3cm}}
\def\secnon#1{\section*{\hspace*{-0.72cm}
 \normalsize\bf$\;$#1}\vspace*{-0.3cm}}
 \def\subsec#1{\addtocounter{subsection}{1}\subsection*{\hspace*{-0.4cm}
 \normalsize\bf\arabic{section}.\arabic{subsection}.$\;$#1}\vspace*{-0.3cm}}
% ------------------------------------------------------------
\vspace*{-1.7cm}
\begin{flushright}
$\vcenter{
%%% \phantom{  \hbox{{\footnotesize OUS and TOKUSHIMA Report}}  }
\hbox{{\footnotesize OUS and TOKUSHIMA Report}}
%%% \phantom{  {\hbox{{\footnotesize TOKUSHIMA Report}}}  }
%{ \hbox{(arXiv:1206.2413)}  }
}$
\end{flushright}

\vskip 1.4cm
\begin{center}
% \hspace*{-0.6cm}
%   {\large\bf Study of possible anomalous top-quark interactions}
  {\large\bf Final charged-lepton angular distribution and possible}
% $\mbox{\large\bf Final charged-lepton angular distributions and possible}$
% Applied to Search for}$

\vskip 0.18cm
%   {\large\bf via the final-lepton angular distribution at LHC}
%  {\large\bf anomalous top-quark couplings in $\mib{t \to \ell^+ X}$ at LHC}
  {\large\bf anomalous top-quark couplings in $\mib{pp\to t \bar{t}X \to \ell^+ X'}$}

\end{center}

\vspace{0.9cm}
\begin{center}
\renewcommand{\thefootnote}{\alph{footnote})}
Zenr\=o HIOKI$^{\:1),\:}$\footnote{E-mail address:
\tt hioki@tokushima-u.ac.jp}\ and\
Kazumasa OHKUMA$^{\:2),\:}$\footnote{E-mail address:
\tt ohkuma@ice.ous.ac.jp}
\end{center}

\vspace*{0.4cm}
\centerline{\sl $1)$ Institute of Theoretical Physics,\ University of Tokushima}

\centerline{\sl Tokushima 770-8502, Japan}

\vskip 0.2cm
\centerline{\sl $2)$ Department of Information and Computer Engineering,}

\centerline{\sl Okayama University of Science}

\centerline{\sl Okayama 700-0005, Japan}

\vspace*{2.3cm}
\centerline{ABSTRACT}

\vspace*{0.2cm}
\baselineskip=21pt plus 0.1pt minus 0.1pt
Possible anomalous (or nonstandard) top-quark interactions with the gluon and
those with the $W$ boson induced by $SU(3) \times SU(2) \times U(1)$ gauge-invariant
dimension-6 effective operators are studied in $pp\to t\bar{t}X \to \ell^+X'$
($\ell=e$ or $\mu$) at the Large Hadron Collider (LHC). The final charged-lepton
($\ell^+$) angular distribution is first computed for nonvanishing nonstandard
top-gluon and top-$W$ couplings with a cut on its transverse momentum.
The optimal-observable procedure is then applied to this distribution in order
to estimate the expected statistical uncertainties in measurements of those
couplings that contribute to this process in the leading order.

\vskip 1.5cm

\vfill
% PACS:  12.38.-t, 12.38.Bx, 12.38.Qk, 12.60.-i, 14.65.Ha, 14.70.Dj

% Keywords:
% Hadron colliders, Anomalous couplings, Top productions,\\ \ \ \ \ \ \  Lepton distributions
\setcounter{page}{0}
\newpage
\renewcommand{\thefootnote}{$\sharp$\arabic{footnote}}
%-------------------------------------------------------------
\pagestyle{plain} \setcounter{footnote}{0}

% 111111111111111111111111111111111111111111111111111111111111
\sec{Introduction}
% 111111111111111111111111111111111111111111111111111111111111

The Large Hadron Collider (LHC) at CERN has been presenting us fruitful experimental
data on various particles/processes ever since it started operating, of course including
the historic discovery of the/a Higgs boson~\cite{LHC}. Exploring possible new physics
beyond the standard model (BSM) is also an important mission of the LHC. Although
they have not found so far any exciting signals indicating BSM yet, this fact never means
that there do not exist exotic particles since their masses might be too high to be
directly produced there.

Even in such a case, we still would be able to investigate certain new-physics effects
indirectly, using data from the LHC. For example, we have studied possible nonstandard
chromomagnetic and chromoelectric dipole moments of the  top-quark (denoted as $d_V$
and $d_A$ respectively) in Refs.\cite{Hioki:2009hm}--\cite{Hioki:2013hva}, and obtained
much stronger restrictions on them than before\footnote{As for the preceding analyses,
    see the reference lists of \cite{Hioki:2009hm}--\cite{Hioki:2013hva}.}\
by adding the data on the $t\bar{t}$ total cross sections from the LHC to those from the
Tevatron. We then carried out an optimal-observable analysis (OOA) to show how precisely
we could determine those nonstandard couplings in $pp \to t\bar{t}X \to \ell^+ X'$ ($\ell=e$
or $\mu$) under a linear
approximation by using the $\ell^+$ angular and energy distributions, where we also took
into account possible nonstandard top-$W$ coupling (denoted as $d_R$) \cite{Hioki:2012vn}.
There, however, we were not able to study the $d_R$ contribution through the angular
distribution due to the decoupling theorem \cite{Grzadkowski:1999iq}--\cite{Godbole:2006tq}.

The $d_R$ dependence of this distribution recovers if we perform the energy integration
necessary to derive it in some limited range, as will be discussed later. The purpose of
this article is to study if we could thereby draw any new information on $d_R$ via a similar
OOA: After summarizing our calculational framework, we are
going to clarify to what extent the distribution becomes dependent of this parameter by
computing it for some different $d_R$ values with a $\ell^+$ transverse-momentum
($p_{\ell\,{\rm T}}$) cut. Then we apply the optimal-observable procedure to this
distribution with and without the $d_V$-term contribution. Concerning the $\ell^+$ energy
distribution, on the other hand, we do not re-study it here because that distribution is
$d_R$-dependent from the beginning and therefore adding the $p_{\ell\,{\rm T}}$ cut does
not bring us anything essentially-new in comparison with what we have done in
\cite{Hioki:2012vn}.

% 2222222222222222222222222222222222222222222222222222222222222222
\sec{Framework}
% 22222222222222222222222222222222222222222222222222222222222222222
%\subsec{Effective Lagrangian}

The framework of our model-independent analyses is based on an effective-La\-gran\-g\-i\-an
whose low-energy form reproduces the standard-model (SM) interactions. This is one of the most
promising methods to describe new-physics phenomena when the energy of our experimental
facility is not high enough to produce new particles. Assuming any non-SM particles too heavy
to appear as real ones, we take the following effective Lagrangian:
\begin{equation}\label{eq:efflag}
{\cal L}_{\rm eff}={\cal L}_{\rm SM} 
+ \frac{1}{{\varLambda}^2} \sum_i \left(\,  {C_i \cal O}_i + {\rm h.c.} \,\right),
\end{equation}
where ${\cal L}_{\rm SM}$ is the SM Lagrangian, ${\cal O}_i$ mean
$SU(3)\times SU(2)\times U(1)$ gauge-invariant operators of mass-dimension 6 involving
only the SM fields and their coefficients $C_i$ parameterize virtual effects of new
particles at an energy less than the assumed new-physics scale ${\varLambda}$. Note here
that the dimension-6 operators give the largest contributions in relevant processes
as long as we assume the lepton-number conservation. In this framework, all the form
factors related to $C_i$ are dealt with as constant parameters, without supposing any
specific new-physics models.

All those dimension-6 operators have been arranged in
Refs.\cite{Buchmuller:1985jz}--\cite{Grzadkowski:2010es}. Following the notation of
\cite{AguilarSaavedra:2008zc}, the effective Lagrangian for the parton-level process
$q \bar{q} /gg  \to t \bar{t} \to b \bar{b} W^+ W^-$ is given in~\cite{HIOKI:2011xx} as
\begin{alignat}{1}\label{eq:efflag_3rd}
 {\cal L}_{\rm eff} &={\cal L}_{t\bar{t}g,gg}+{\cal L}_{tbW}  \\
 &  {\cal L}_{t\bar{t}g,gg}
    = -\frac{1}{2} g_s \sum_a \Bigl[\,\bar{\psi}_t(x)\lambda^a \gamma^{\mu}
    \psi_t(x) G_\mu^a(x)
    \bigl. \nonumber\\
 &\phantom{====}
    -\bar{\psi}_t(x)\lambda^a\frac{\sigma^{\mu\nu}}{m_t}\bigl(d_V+id_A\gamma_5\bigr)
  \psi_t(x)G_{\mu\nu}^a(x)\,\Bigr], \\
 &  {\cal L}_{tbW}  = -\frac{1}{\sqrt{2}}g 
  \Bigl[\,\bar{\psi}_b(x)\gamma^\mu(f_1^L P_L + f_1^R P_R)\psi_t(x)W^-_\mu(x) \Bigr.
  \nonumber\\
 &\phantom{====}+\bar{\psi}_b(x)\frac{\sigma^{\mu\nu}}{M_W}(f_2^L P_L + f_2^R P_R)
   \psi_t(x)\partial_\mu W^-_\nu(x) \,\Bigr], 
\end{alignat}
where $g_s$ and $g$ are the $SU(3)$ and $SU(2)$ coupling constants, 
$P_{L/R}\equiv(1\mp\gamma_5)/2$,
$d_V, d_A$ and $f_{1,2}^{L,R}$ are form factors defined as
\begin{alignat}{2}\label{eq:dvdadef}
 d_V &\equiv \frac{\sqrt{2}v m_t}{g_s {\varLambda}^2} {\rm Re}(C^{33}_{uG\phi}),
 & \quad d_A&\equiv \frac{\sqrt{2}v m_t}{g_s {\varLambda}^2} {\rm Im}(C^{33}_{uG\phi}),
   \nonumber\\
  f_1^L&\equiv V_{tb}+C^{(3,33)*}_{\phi q}\frac{v^2}{{\varLambda}^2},  
 & \quad  f_1^R&\equiv C^{33*}_{\phi \phi}\frac{v^2}{2{\varLambda}^2},  \\
  f_2^L&\equiv -\sqrt{2} C^{33*}_{dW}\frac{v^2}{{\varLambda}^2},
 & \quad f_2^R&\equiv -\sqrt{2} C^{33}_{uW}\frac{v^2}{{\varLambda}^2} \nonumber
\end{alignat}
with $v$ being the Higgs vacuum expectation value and $V_{tb}$ being the ($tb$) element
of Kobayashi--Maskawa matrix. Among those unknown parameters, $d_V$ and $d_A$ are
respectively the top-quark chromo\-magnetic and chromo\-electric dipole moments,
and we use $d_R$ defined as 
\begin{equation}
d_R\equiv {\rm Re}(f_2^R) M_W /m_t \label{eq:drdef}
\end{equation}
instead of $f_2^R$ in order to make our formulas a little bit simpler.

In the following work, we use the above effective Lagrangian for top-quark
interactions, and adopt the linear approximation for those nonstandard parameters
as in \cite{Hioki:2012vn}, where $d_V$ and $d_R$ come into our analyses (note that
$d_A$ terms do not contribute to $q\bar{q}/gg \to t\bar{t}$ in the leading order
because of their $CP$-odd property).
We assume the other interactions, e.g. the one for $W^+ \to \ell^+ \nu$, are described
by the usual SM Lagrangian, and all the fermions lighter than the top quark are treated
as massless particles. Concerning the parton distribution functions, we have
been using CTEQ6.6M (NNLO approximation) \cite{Nadolsky:2008zw}.

% 333333333333333333333333333333333333333333333333333333333333
\sec{Lepton angular distribution and decoupling theorem}
% 333333333333333333333333333333333333333333333333333333333333

What we call ``the decoupling theorem'' is a theorem which states that the leading
contribution of the anomalous top-decay couplings, $d_R$ in our case, to final-particle
angular distributions vanishes when only a few conditions are satisfied
\cite{Grzadkowski:1999iq}--\cite{Godbole:2006tq}. In terms of the $\ell^+$ angular
distribution under consideration, this theorem holds if we assume the standard $V-A$
structure for the $\nu\ell W$ coupling and perform the lepton-energy integration fully
over the kinematically-allowed range. As a result, this distribution becomes exclusively
dependent of $d_V$. That is, we can no longer get any information thereby on the nonstandard
top-decay coupling $d_R$.

Although it is not possible to cover the full phase space of
the final-lepton momentum in actual experiments, we could carry out the above energy
integration using the energy distribution reconstructed through a proper extrapolation.
Therefore the above-mentioned full integration is not unrealistic. This however tells
us that we might be able to draw certain new information on $d_R$ by using the angular
distribution with some cut on the lepton momentum.

Let us calculate the $\ell^+$ angular distribution with a $\ell^+$ transverse-momentum
($p_{\ell\,{\rm T}}$) cut as a typical and realistic experimental condition. We first
take one of the proton beams as the base axis and express the differential cross section
of $pp\to t\bar{t} X \to \ell^+ X'$ (the angular and energy distribution of $\ell^+$)
in the proton-proton CM frame as follows:
\begin{equation}
\frac{d^2\sigma_{\ell}}{d E_{\ell}\,d \cos\theta_\ell} 
  =  f_{\rm SM}(E_{\ell}, \cos\theta_\ell) + d_V f_{d_V}(E_{\ell},\cos\theta_\ell) 
  + d_R f_{d_R}(E_{\ell},\cos\theta_\ell),
\end{equation}
where $E_{\ell}$ is the lepton energy, $\theta_\ell$ is the lepton scattering angle, i.e.,
the angle formed by the $\ell^+$ momentum and
the above-mentioned base axis, $f_{\rm SM}(E_{\ell}, \cos\theta_\ell)$ denotes the SM
contribution, and the
other two $f_I(E_{\ell}, \cos\theta_\ell)$ describe the non-SM terms corresponding to their
coefficients. The explicit forms of $f_I(E_{\ell}, \cos\theta_\ell)$ at the parton level
are easily found in the relevant formulas in~\cite{HIOKI:2011xx}.
Then, the $\ell^+$ angular distribution is written as
\begin{equation}
 \frac{d\sigma_{\ell}}{d \cos\theta_{\ell}} 
 =  g_1(\cos\theta_\ell) + d_V ~g_2(\cos\theta_\ell) + d_R~g_3(\cos\theta_\ell),
\label{g123}
\end{equation} 
where $g_i(\cos\theta_\ell)$ are given by
\begin{equation}
  g_i(\cos\theta_\ell) = \int d E_\ell~f_I(E_\ell,\cos\theta_\ell)
\label{g123def}
\end{equation}
with $i=1,2$ and 3 corresponding to $I={\rm SM}$, $d_V$ and $d_R$, respectively.
In the above $E_\ell$ integration, the kinematically-allowed range is
\begin{equation}
\frac{M_W^2}{\sqrt{s}(1+\beta)} \leq E_\ell \leq \frac{m_t^2}{\sqrt{s}(1-\beta)}
\label{Elrange1}
\end{equation}
with $\beta \equiv \sqrt{1-4m_t^2/s}$. As mentioned, $g_3(\cos\theta_\ell)$ disappears
if we perform the integration fully over this range due to the decoupling theorem.

We compute this angular distribution for $\sqrt{s}=$ 14 TeV\footnote{We performed analyses
    for $\sqrt{s}=$ 7, 8, 10 and 14 TeV in \cite{Hioki:2012vn}, but we here focus on 14 TeV
    since the LHC is now being upgraded toward this energy.}\ 
and $p_{\ell\,{\rm T}} \geq p_{\ell\,{\rm T}}^{min}$, the latter of which leads to
the lower bound of $E_\ell$ as
\begin{equation}
E_\ell \geq p_{\ell\,{\rm T}}^{min}/\sqrt{1-\cos^2 \theta_\ell},
\label{Elrange2}
\end{equation}
and Eqs.(\ref{Elrange1},\ref{Elrange2}) require
\begin{equation}
| \cos\theta_\ell | \leq \sqrt{1-s(1-\beta)^2(p_{\ell\,{\rm T}}^{min}/m_t^2)^2}.
\label{clrange}
\end{equation}
Practically, however, this restriction on $\cos\theta_\ell$ affects its range only a little,
e.g., the right-hand side of this inequality is 0.999898 even for
$p_{\ell\,{\rm T}}^{min}=$ 100 GeV.

\begin{figure}[H]
\begin{minipage}{14cm}
\begin{center}
\vspace*{0.cm} \hspace*{-0cm}
% \psfrag{dR}{$d_R$}
\includegraphics[width=10.8cm]{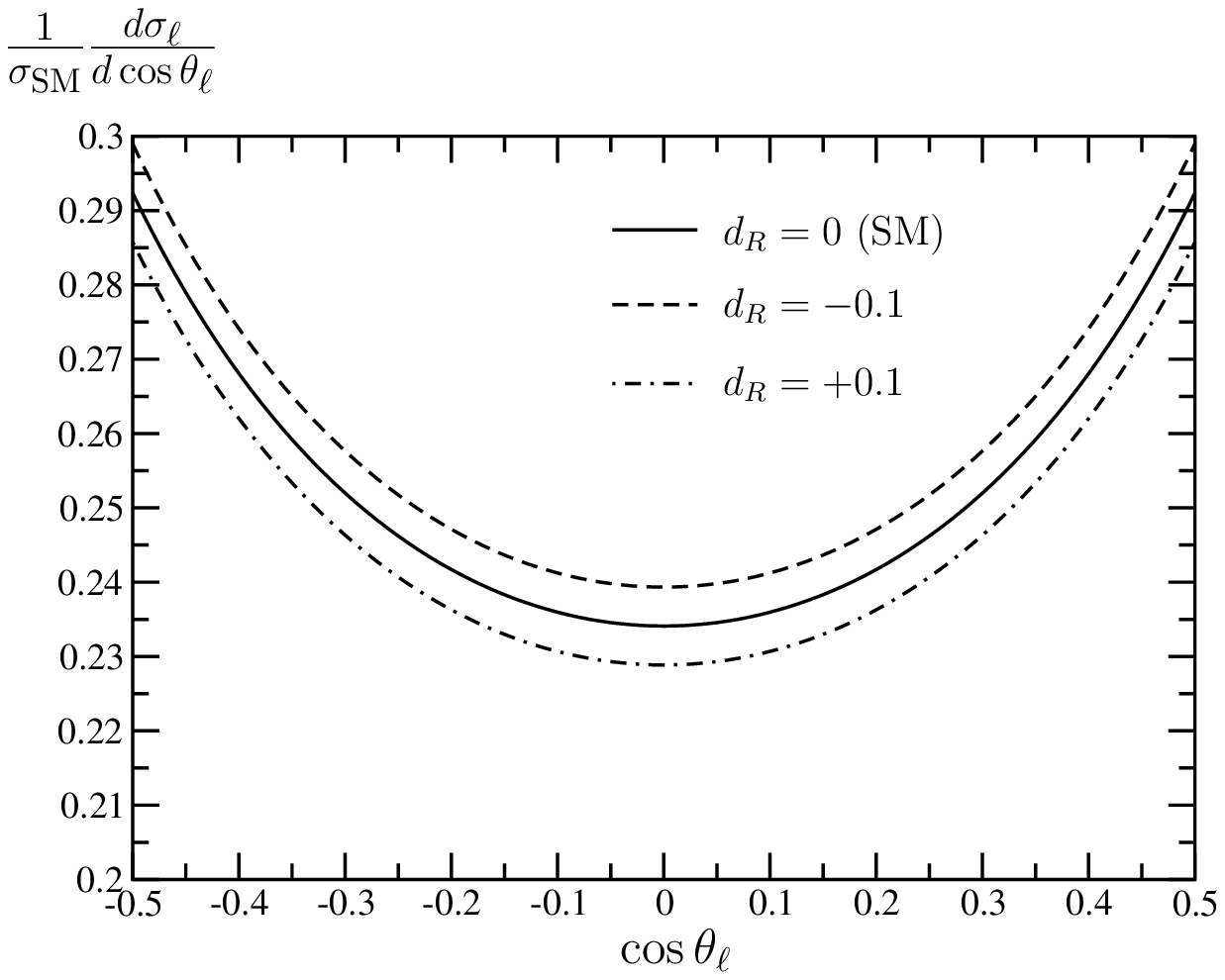}
\vspace*{0.1cm}
\caption{The $\ell^+$ angular distributions (without the $d_V$ terms and
normalized by the SM total cross section with no $p_{\ell\,{\rm T}}$ constraint)
for $p_{\ell\,{\rm T}}^{min}=20$ GeV, and $d_R=0$ (SM), $-0.1$ and $+0.1$.}
\label{angular1}
\end{center}
\end{minipage}
\end{figure}

\begin{figure}[H]
\begin{minipage}{14cm}
\begin{center}
% \vspace*{1.5cm} \hspace*{-2.45cm}
\includegraphics[width=10.8cm]{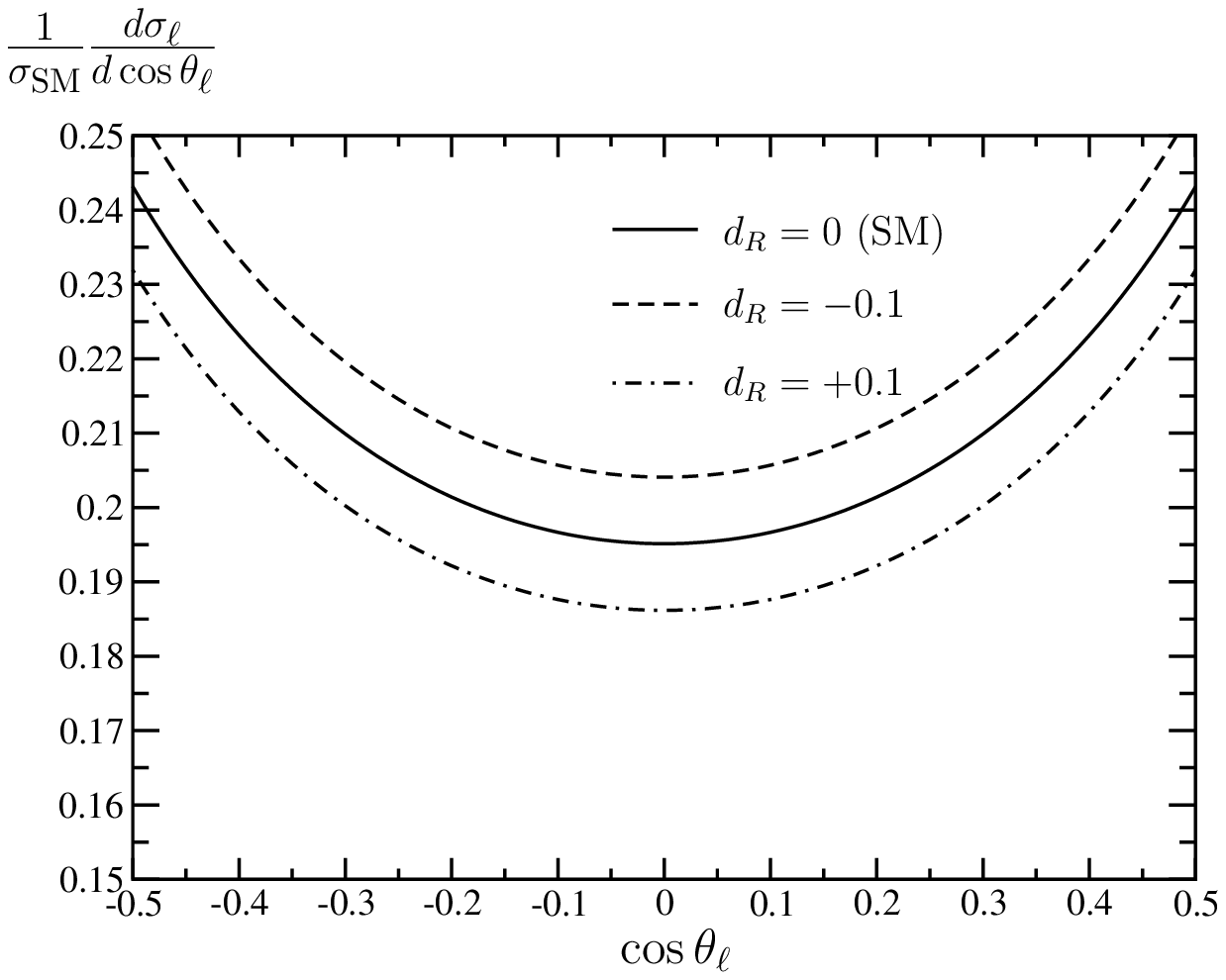}
\vspace*{0.1cm}
\caption{The $\ell^+$ angular distributions (without the $d_V$ terms and
normalized by the SM total cross section with no $p_{\ell\,{\rm T}}$ constraint)
for $p_{\ell\,{\rm T}}^{min}=30$ GeV, and $d_R=0$ (SM), $-0.1$ and $+0.1$.}
\label{angular2}
\end{center}
\end{minipage}
\end{figure}

\begin{figure}[H]
\begin{minipage}{14cm}
\begin{center}
%\vspace*{0.5cm} \hspace*{-2.45cm}
\includegraphics[width=10.8cm]{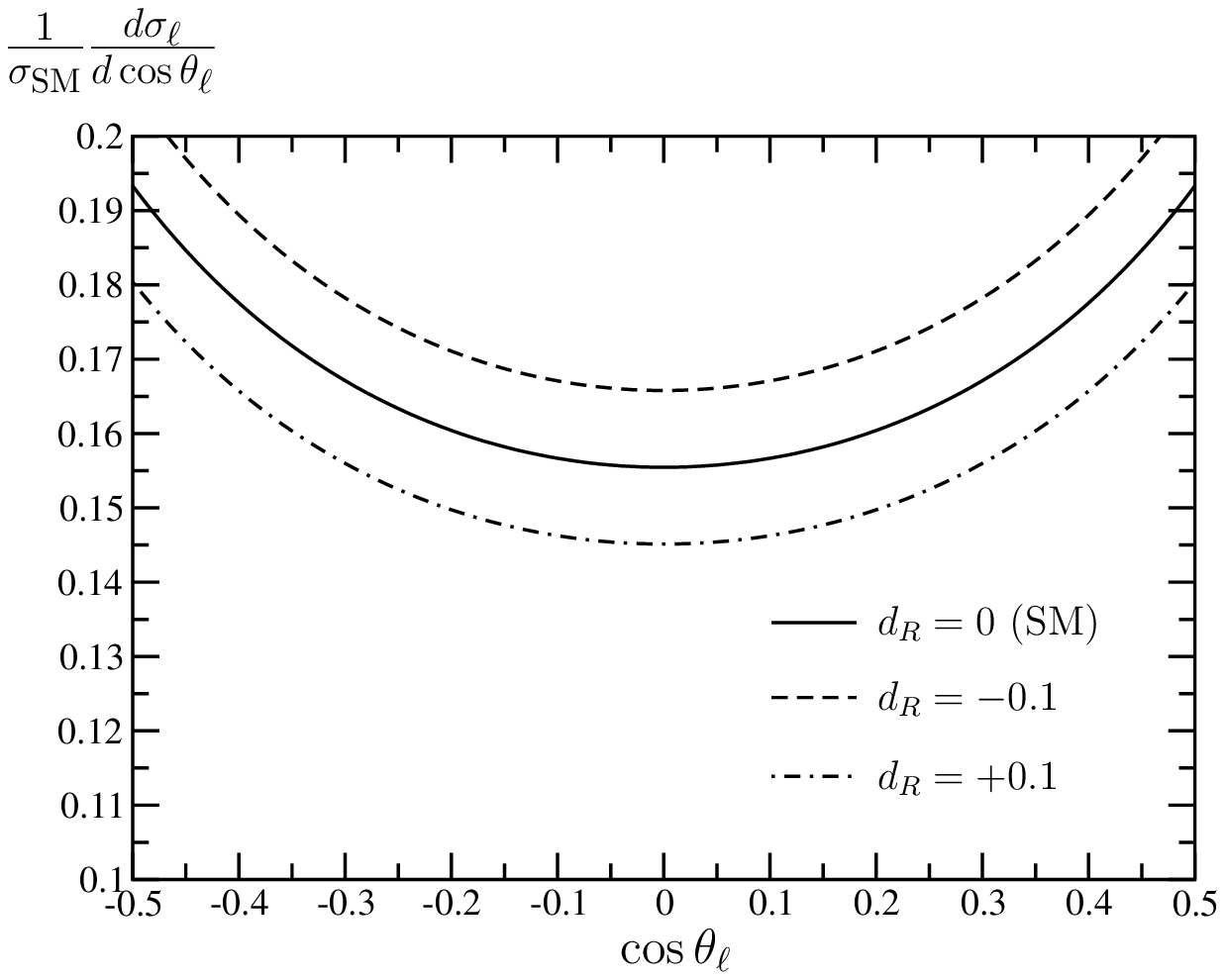}
\vspace*{0.1cm}
\caption{The $\ell^+$ angular distributions (without the $d_V$ terms and
normalized by the SM total cross section with no $p_{\ell\,{\rm T}}$ constraint)
for $p_{\ell\,{\rm T}}^{min}=40$ GeV, and $d_R=0$ (SM), $-0.1$ and $+0.1$.}
\label{angular3}
\end{center}
\end{minipage}
\end{figure}

\begin{figure}[H]
\begin{minipage}{14cm}
\begin{center}
% \vspace*{0.5cm} \hspace*{-3.0cm}
%\psfrag{dR}{$d_R$}
\includegraphics[width=10.8cm]{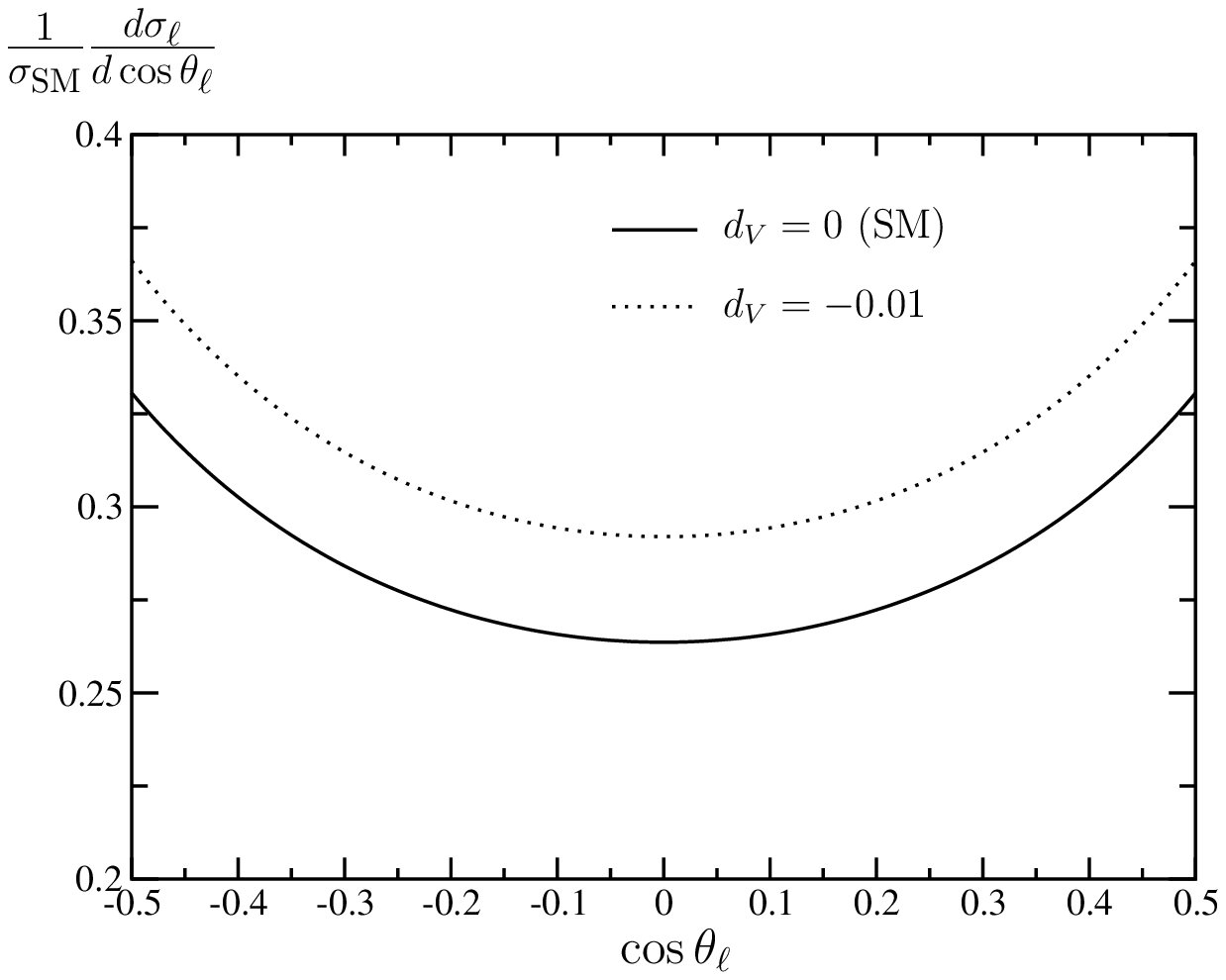}
\vspace*{0.1cm}
\caption{The $\ell^+$ angular distributions (without the $d_R$ terms and
normalized by the SM total cross section with no $p_{\ell\,{\rm T}}$ constraint)
for $d_V=0$ (SM) and $-0.01$. We did not impose any $p_{\rm T}$ cut here
because the $d_V$ effects are free from the decoupling theorem.}
\label{angular4}
\end{center}
\end{minipage}
\end{figure}

We show the $d_R$ dependence of the angular distribution within the range $|d_R| \leq 0.1$
\cite{Grzadkowski:2008mf,Prasath:2014mfa} in Figs.\ref{angular1}--\ref{angular3}, where
we normalized the distribution by the SM total cross section of the same process but with
no $p_{\ell\,{\rm T}}$ constraint: $\sigma_{\rm SM}=134$ pb, and varied the cut as
$p_{\ell\,{\rm T}}^{min}=20,\,30,\,40$ GeV for $m_t=$ 173 GeV. As for $d_V$ we simply set
it equal to zero there since what we are interested in is the $d_R$ dependence. Then we
show similar curves but for $d_V=-0.01$ \cite{Hioki:2013hva} (with $d_R=0$ and
no $p_{\ell\,{\rm T}}$ cut) in Fig.\ref{angular4} for comparison.
In all the Figures, we limit the horizontal range to $|\cos \theta_{\ell}| \leq 0.5$
simply because the $d_{V,R}$ effects become less clear if we draw the curves over
the full range given by Eq.(\ref{clrange}).

\vskip 1.cm

%%%%%%%%%%%%%%%%%%%%%%%%%%%%%%%%%%%%%%%%%%%%%%%%%%%%%%%%%%%%%%%%%%%%%
%%%%%%%%%%%%%%%%%%%%%%%%%%%%%%%%%%%%%%%%%%%%%%%%%%%%%%%%%%%%%%%%%%%%%%%%%

\newpage

%%%%%%%%%%%%%%%%%%%%%%%%%%%%%%%%%%%%%%%%%%%%%%%%%%%%%%%%%%%%%%%%%%%%%
%%%%%%%%%%%%%%%%%%%%%%%%%%%%%%%%%%%%%%%%%%%%%%%%%%%%%%%%%%%%%%%%%%%%%%%%%

\vskip 0.5cm

%%%%%%%%%%%%%%%%%%%%%%%%%%%%%%%%%%%%%%%%%%%%%%%%%%%%%%%%%%%%%%%%%%%%%
%%%%%%%%%%%%%%%%%%%%%%%%%%%%%%%%%%%%%%%%%%%%%%%%%%%%%%%%%%%%%%%%%%%%%%%%%

\newpage

%%%%%%%%%%%%%%%%%%%%%%%%%%%%%%%%%%%%%%%%%%%%%%%%%%%%%%%%%%%%%%%%%%%%%
%%%%%%%%%%%%%%%%%%%%%%%%%%%%%%%%%%%%%%%%%%%%%%%%%%%%%%%%%%%%%%%%%%%%%%%%%

\vspace*{0.8cm}

We see through Figs.\ref{angular1}--\ref{angular3} that the angular distribution
with a $p_{\ell\,{\rm T}}$ cut has actually become $d_R$ dependent although
it is not as large as the $d_V$ contribution in Fig.\ref{angular4}. In order to
show these ${\cal O}(d_R)$ corrections to the SM distributions (with the same
$p_{\ell\,{\rm T}}$ cut) more quantitatively, let us present their sizes at
$\cos\theta_{\ell}=0$ for $d_R=0.1$ as an example:
\begin{equation}
p_{\ell\,{\rm T}}^{min}=20\ {\rm GeV}:\ -2.2\:\%,\ \ \ \
30\ {\rm GeV}:\ -4.6\:\%,\ \ \ \ 40\ {\rm GeV}:\ -6.6\:\%. 
\end{equation}

% 444444444444444444444444444444444444444444444444444444444444
\sec{Optimal-observable analysis with ${\mib p_{\ell\,{\rm\bf T}}}$ cut}
% 444444444444444444444444444444444444444444444444444444444444

The optimal-observable analysis (OOA) is a way that could systematically estimate
the expected statistical uncertainties of measurable parameters. Here we apply
this procedure to the $\ell^+$ angular distribution studied in the preceding section.

Leaving its detailed and specific description to \cite{Atwood:1991ka}--\cite{Gunion:1996vv},
let us show how to compute the uncertainties thereby: \\
What we have to do first is to calculate the following $3 \times 3$ matrix
\begin{equation}
M_{ij}^c \equiv \int d\cos\theta_{\ell} 
\frac{g_i(\cos\theta_\ell) g_j(\cos\theta_\ell)}
     {g_1(\cos\theta_\ell)}\ \ \ \ (i,\,j=1,\,2,\,3)
\end{equation}
using $g_{1,2,3}$ defined in Eqs.(\ref{g123},\ref{g123def}), and next its inverse matrix
$X_{ij}^{c}$, both of which are apparently symmetric.\footnote{In our
    preceding OOA \cite{Hioki:2012vn}, we distinguished those quantities computed from
    the angular and energy distributions by adding them superscripts ``$c$'' and ``$E$''
    respectively. Here we do not need such a superscript but we left it for easy comparison
    with our previous results.}
This integration is to be performed over the range given by Eq.(\ref{clrange}).
Then the statistical uncertainties for the measurements of couplings
$d_V$ and $d_R$ could be estimated by
\begin{eqnarray}
&&| \delta d_V |= \sqrt{X_{22}^{c}\sigma_{\ell}/N_{\ell}}
=\sqrt{X_{22}^{c}/L}\,, \label{eq:uncertainty1}\\
&&| \delta d_R |= \sqrt{X_{33}^{c}\sigma_{\ell}/N_{\ell}}
=\sqrt{X_{33}^{c}/L}\label{eq:uncertainty2},
\end{eqnarray}
where $\sigma_{\ell}$, $N_{\ell}$ and $L$ denote the total cross section, the number of
events and the integrated luminosity for the process $pp \to t\bar{t}X \to \ell^+ X'$,
respectively.

We are now ready to carry out necessary numerical computations. Below we show the
elements of $M^{c}$ computed for $\sqrt{s}=$14 TeV:
\\
(1) $p_{\ell\,{\rm T}}^{min}=$ 20 GeV
\begin{equation}
 \begin{array}{lll}
  M^c_{11}=+113.30234,  & M^c_{12}=-1207.01858,& M^c_{13}=-28.89719,\\
  M^c_{22}=+12861.00330,& M^c_{23}=+306.88261, & M^c_{33}=+7.81915.
  \end{array} \label{M1120}
\end{equation}
(2) $p_{\ell\,{\rm T}}^{min}=$ 30 GeV
\begin{equation}
 \begin{array}{lll}
  M^c_{11}=+92.40192,   & M^c_{12}=-982.35568, & M^c_{13}=-45.55635,\\
  M^c_{22}=+10446.10110,& M^c_{23}=+483.46228, & M^c_{33}=+22.82778.
  \end{array} \label{M1130}
\end{equation}
(3) $p_{\ell\,{\rm T}}^{min}=$ 40 GeV
\begin{equation}
 \begin{array}{lll}
  M^c_{11}=+72.29773,  & M^c_{12}=-766.31899, & M^c_{13}=-50.54475,\\
  M^c_{22}=+8124.55885,& M^c_{23}=+535.07185, & M^c_{33}=+35.59435.
  \end{array} \label{M1140}
\end{equation}
Here all these results were derived from the cross section in [pb] unit. Using
the inverse matrices calculated from these elements, we can estimate the statistical
uncertainties of the relevant couplings $\delta d_V$ and $\delta d_R$ according to
Eqs.(\ref{eq:uncertainty1},\ref{eq:uncertainty2}) (Two-parameter analysis).

The set of $M_{ij}^c$ (\ref{M1120})--(\ref{M1140}) also enables us to give another
numerical results. That is, we can do a similar analysis but assuming only $d_R$ is
unknown. This assumption is never unreasonable because we already have shown that
we would be able to obtain good information on $d_V$ (and $d_A$) through the total
cross section of $pp/p\bar{p} \to t\bar {t}X$
without being affected by the top-decay processes. All we have to do for that is perform
the same computations but without the $d_V$ component, i.e., compute the $2\times 2$
matrix $X_{ij}^c$ from $M^c_{ij}$ with $i,j=1,3$, and use Eq.(\ref{eq:uncertainty2})
(One-parameter analysis).

Before giving the results, however, we should remember that we encountered an instability
problem when computing the inverse-matrix in our previous analysis \cite{Hioki:2012vn}.
That is, the numerical results fluctuated
to a certain extent (beyond our expectation) depending on to which decimal places of $M^{c,E}$
we take into account as our input data. Therefore we compute here $X^c_{ii}$ not only for
the above $M^c_{ij}$ but also for those to three and one decimal places in order to check
to what extent the results are stable:
\\
$\bullet$ Two-parameter analysis
\\
(1) $p_{\ell\,{\rm T}}^{min}=$ 20 GeV
\begin{equation}
X^c_{22}= 2.1\,(2.3,\ 2.2), \ \ \ X^c_{33}= 11.7\,(12.6,\ 14.0).
\end{equation}
(2) $p_{\ell\,{\rm T}}^{min}=$ 30 GeV
\begin{equation}
X^c_{22}= 3.1\,(2.9,\ *\!*\!*),\ \ \ X^c_{33}= 19.7\,(18.4,\ *\!*\!*).
\end{equation}
(3) $p_{\ell\,{\rm T}}^{min}=$ 40 GeV
\begin{equation}
X^c_{22}= 5.2\,(4.7,\ 0.4),\ \ \ X^c_{33}= 40.0\,(36.2,\ 3.1).
\end{equation}
Here all the figures in the parentheses are from $M^c_{ij}$ rounded off properly to three and
one decimal places respectively, and $*\!*\!*$ expresses that we had no meaningful solutions
there, i.e., the results became negative. The results for $p_{\ell\,{\rm T}}^{min}=$ 20 GeV
seem to be rather stable, but there is non-negligible instability in the results for
$p_{\ell\,{\rm T}}^{min}=$ 30 and 40 GeV. Therefore we conclude that we would not obtain reliable
results from the two-parameter analysis unless the corresponding cross sections are determined
very precisely, i.e., at least to three-decimal-place precision. As discussed in \cite{Hioki:2012vn},
this problem would come from dominant $d_V$-term contributions. Indeed, the following results of
the one-parameter analysis without the $d_V$ term are quite stable. 
\\
$\bullet$ One-parameter analysis
\\
(1) $p_{\ell\,{\rm T}}^{min}=$ 20 GeV
\begin{equation}
X^c_{33}= 2.2\,(2.2,\ 2.3).
\end{equation}
(2) $p_{\ell\,{\rm T}}^{min}=$ 30 GeV
\begin{equation}
X^c_{33}= 2.7\,(2.7,\ 3.4).
\end{equation}
(3) $p_{\ell\,{\rm T}}^{min}=$ 40 GeV
\begin{equation}
X^c_{33}= 3.9\,(3.9,\ 3.1).
\end{equation}
These results present us a hint for anomalous-top-couplings search through the lepton
angular distribution, i.e., it will be effective (and also inevitable) to combine its
data with those of the total $t\bar{t}$ cross section where we will be able to explore
$d_V$ (and also $d_A$) in detail.

Let us present our final results for the one-parameter analysis. The expected statistical
uncertainties in measuring $d_R$ are estimated as follows:
\\
(1) $p_{\ell\,{\rm T}}^{min}=$ 20 GeV
\begin{equation}
| \delta d_R | = (1.5 \pm 0.0)/\sqrt{L}.
\end{equation}
(2) $p_{\ell\,{\rm T}}^{min}=$ 30 GeV
\begin{equation}
| \delta d_R | = (1.7 \pm 0.1)/\sqrt{L}.
\end{equation}
(3) $p_{\ell\,{\rm T}}^{min}=$ 40 GeV
\begin{equation}
| \delta d_R | = (1.9 \pm 0.1)/\sqrt{L}.
\end{equation}
For instance, if $L=1000\ {\rm pb}^{-1}$ is achieved and if there exists nonstandard $d_R$
coupling with the size $d_R=0.1$, we would be able to confirm its effects at $2.1\sigma$
level (apart from the systematic errors)
via an analysis using $p_{\ell\,{\rm T}}^{min}=$ 20 GeV.

Finally, before closing this section, another comment would be also necessary on QCD
higher-order corrections since all the numerical computations here were done with the tree-level
formulas. In order to take into account those corrections, we multiply the tree cross sections
by the $K$-factor ($K \simeq 1.5$ \cite{Kidonakis:2013zqa}). This factor disappears in the
combination $X_{ii}^{c}\sigma_{\ell}$ and remains only in $N_{\ell}\,(=L\sigma_{\ell})$
when we estimate $\delta d_{V,R}$. Therefore the luminosity $L$ in our results should be
understood as an effective one including $K$ (and also the final charged-lepton
detection-efficiency $\epsilon_\ell$).

% 555555555555555555555555555555555555555555555555555555555555
\sec{Summary}
% 555555555555555555555555555555555555555555555555555555555555

We studied possible nonstandard top-gluon and top-$W$ couplings for hadron-collider
experiments through the angular distribution of the final charged-lepton from a
top-quark semileptonic decay in a model-independent way. Those couplings are derived
as parameters which characterize the effects of dimension-6 effective operators based
on the scenario of Buchm\"{u}ller and Wyler~\cite{Buchmuller:1985jz}. More specifically,
we analyzed the top-gluon coupling (denoted as $d_V$) and the top-$W$ coupling (denoted
as $d_R$) which contribute to top-quark pair productions and decays respectively
in the linear approximation as nonstandard interactions.

We are not able to observe the $d_R$-term contribution through the lepton angular distribution
due to the decoupling theorem \cite{Grzadkowski:1999iq}--\cite{Godbole:2006tq}, if we perform
the lepton-energy integration fully over the kinematically-allowed range when deriving
this distribution. Our main purpose here was to explore if we could draw any new
information on $d_R$ via an optimal-observable analysis of this distribution by
introducing a lepton transverse-momentum cut and giving the angular distribution
some $d_R$ dependence.

We found that the distribution thereby becomes actually $d_R$ dependent, which enabled us
to carry out an optimal-observable analysis including the $d_R$-terms, although we encountered
an instability problem in calculating necessary inverse-matrices as in our preceding study
\cite{Hioki:2012vn}. Therefore we will be able to obtain some new information on this parameter.
In fact, this $p_{\ell\,{\rm T}}$ constraint makes the corresponding cross section smaller and
consequently the precision becomes a bit lower than the case of the analysis using the lepton
energy distribution \cite{Hioki:2012vn}. However we still would like to stress that the analyses
here are useful since we should combine all available data in order to explore possible new
physics beyond the standard model.

%%%%%%%%%%%%%%%%%%%%%%%%%%%%%%%%%%%%%%%%%%%%%%%%%%%%%%%%%%%%%%%%%%%%%%%%%%
%
\secnon{Acknowledgments}
%
%%%%%%%%%%%%%%%%%%%%%%%%%%%%%%%%%%%%%%%%%%%%%%%%%%%%%%%%%%%%%%%%%%%%%%%%%%%%%%
This work was partly supported by the Grant-in-Aid for Scientific Research 
No. 22540284 from the Japan Society for the Promotion of Science.
Part of the algebraic and numerical calculations were carried out on the computer
system at Yukawa Institute for Theoretical Physics (YITP), Kyoto University.
\baselineskip=20pt plus 0.1pt minus 0.1pt

 \vspace*{0.8cm}
% RRRRRRRRRRRRRRRRRRRRRRRRRRRRRRRRRRR


\begin{thebibliography}{99}
%
\bibitem{LHC} LHC website: {\tt http://public.web.cern.ch/public/en/LHC/LHC-en.html}
%
\bibitem{Hioki:2009hm}
  Z.~Hioki and K.~Ohkuma,
  %``Search for anomalous top-gluon couplings at LHC revisited,''
  Eur.\ Phys.\ J.\  C {\bf 65} (2010) 127 (arXiv:0910.3049 [hep-ph]);
  %%CITATION = EPHJA,C65,127;%%
%\bibitem{Hioki:2010zu}
  %Z.~Hioki and K.~Ohkuma,
  %``Addendum to: Search for anomalous top-gluon couplings at LHC revisited,''
  {\it ibid.} % Eur.\ Phys.\ J.\  
  C {\bf 71} (2011) 1535 (arXiv:1011.2655 [hep-ph]).
  %%CITATION = ARXIV:1011.2655;%%
\bibitem{HIOKI:2011xx}
  Z.~Hioki and K.~Ohkuma,
  %``Exploring anomalous top interactions via the final lepton in $t\bar{t}$
  % productions/decays at hadron colliders,''
  Phys.\ Rev.\ D {\bf 83} (2011) 114045
  (arXiv:1104.1221 [hep-ph]); %.
  %%CITATION = ARXIV:1104.1221;%%
\bibitem{Hioki:2013hva}
  Z.~Hioki and K.~Ohkuma,
  %``Latest constraint on nonstandard top-gluon couplings at hadron colliders and
  % its future prospect,''
  Phys.\ Rev.\ D {\bf 88} (2013) 017503
  (arXiv:1306.5387 [hep-ph]).
  %%CITATION = ARXIV:1306.5387;%%
\bibitem{Hioki:2012vn}
  Z.~Hioki and K.~Ohkuma,
  %``Optimal-observable Analysis of Possible Non-standard Top-quark Couplings
  % in $pp -> t \bar{t} X -> l^+ X'$,''
  Phys.\ Lett.\ B {\bf 716} (2012) 310
  (arXiv:1206.2413 [hep-ph]).
  %%CITATION = ARXIV:1206.2413;%%
\bibitem{Grzadkowski:1999iq}
  B.~Grzadkowski and Z.~Hioki,
  %``New hints for testing anomalous top quark interactions at future linear
  %colliders,''
  Phys.\ Lett.\  B {\bf 476} (2000) 87 (hep-ph/9911505);
  %%CITATION = PHLTA,B476,87;%%
%\bibitem{Grzadkowski:2001tq}
  %B.~Grzadkowski and Z.~Hioki,
  %``Angular distribution of leptons in general $t \bar{t}$ production and
  %decay,''
  {\it ibid.} % Phys.\ Lett.\  
  B {\bf 529} (2002) 82 (hep-ph/0112361);
  %%CITATION = PHLTA,B529,82;%%
% \bibitem{Grzadkowski:2002gt}
%  B.~Grzadkowski and Z.~Hioki,
  %``Decoupling of anomalous top-decay vertices in angular distribution of
  %secondary particles,''
  {\it ibid.} % Phys.\ Lett.\
  B {\bf 557} (2003) 55 (hep-ph/0208079).
  %%CITATION = PHLTA,B557,55;%%
\bibitem{Rindani:2000jg}
  S.D.~Rindani,
  %``Effect of anomalous t b W vertex on decay-lepton distributions in  e+ e-
  %--> t anti-t and CP-violating asymmetries,''
  Pramana {\bf 54} (2000) 791 (hep-ph/0002006).
  %%CITATION = PRAMC,54,791;%%
\bibitem{Godbole:2006tq}
  R.M.~Godbole, S.D.~Rindani and R.K.~Singh,
  %``Lepton distribution as a probe of new physics in production and decay of
  %the t quark and its polarization,''
  JHEP {\bf 0612} (2006) 021 (hep-ph/0605100).
  %%CITATION = JHEPA,0612,021;%%
\bibitem{Buchmuller:1985jz}
  W.~Buchmuller and D.~Wyler,
  %``Effective Lagrangian Analysis Of New Interactions And Flavor
  %Conservation,''
  Nucl.\ Phys.\  B {\bf 268} (1986) 621.
  %%CITATION = NUPHA,B268,621;%%
\bibitem{Arzt:1994gp}
  C.~Arzt, M.B.~Einhorn and J.~Wudka,
  %``Patterns of deviation from the standard model,''
  Nucl.\ Phys.\  B {\bf 433} (1995) 41
  (hep-ph/9405214).
  %%CITATION = NUPHA,B433,41;%%
\bibitem{AguilarSaavedra:2008zc}
  J.A.~Aguilar-Saavedra,
  % {\it A minimal set of top anomalous couplings},
  Nucl.\ Phys.\  B {\bf 812} (2009) 181 (arXiv:0811.3842 [hep-ph]);
  %%CITATION = NUPHA,B812,181;%%
  %``A minimal set of top-Higgs anomalous couplings,''
  {\it ibid.} % Nucl.\ Phys.\
  B {\bf 821} (2009) 215 (arXiv:0904.2387 [hep-ph]).
  %%CITATION = NUPHA,B821,215;%%
\bibitem{Grzadkowski:2010es}
  B.~Grzadkowski, M.~Iskrzynski, M.~Misiak and J.~Rosiek,
  %``Dimension-Six Terms in the Standard Model Lagrangian,''
  JHEP {\bf 1010} (2010) 085 (arXiv:1008.4884 [hep-ph]).
  %%CITATION = JHEPA,1010,085;%%
\bibitem{Nadolsky:2008zw}
  P.M.~Nadolsky, H.-L.~Lai, Q.-H.~Cao, J.~Huston, J.~Pumplin, D.~Stump,
  W.-K.~Tung and C.-P.~Yuan,
  %``Implications of CTEQ global analysis for collider observables,''
  Phys.\ Rev.\ D {\bf 78} (2008) 013004 (arXiv:0802.0007 [hep-ph]).
  %%CITATION = ARXIV:0802.0007;%%
\bibitem{Grzadkowski:2008mf}
  B.~Grzadkowski and M.~Misiak,
  %``Anomalous Wtb coupling effects in the weak radiative B-meson decay,''
  Phys.\ Rev.\ D {\bf 78} (2008) 077501 [Erratum-ibid.\ D {\bf 84} (2011) 059903]
  (arXiv:0802.1413 [hep-ph]).
  %%CITATION = ARXIV:0802.1413;%%
\bibitem{Prasath:2014mfa}
  A.~Prasath, R.M.~Godbole and S.D.~Rindani,
  %``Top polarisation measurement and anomalous $Wtb$ coupling,''
  arXiv:1405.1264 [hep-ph].
  %%CITATION = ARXIV:1405.1264;%%
\bibitem{Atwood:1991ka}
  D.~Atwood and A.~Soni,
  %``Analysis for magnetic moment and electric dipole moment form-factors
  % of the top quark via e+ e- ---> t anti-t,''
  Phys.\ Rev.\ D {\bf 45} (1992) 2405.
  %%CITATION = PHRVA,D45,2405;%%
\bibitem{Davier:1992nw}
  M.~Davier, L.~Duflot, F.~Le Diberder and A.~Rouge,
  %``The Optimal method for the measurement of tau polarization,''
  Phys.\ Lett.\ B {\bf 306} (1993) 411.
  %%CITATION = PHLTA,B306,411;%%
\bibitem{Diehl:1993br}
  M.~Diehl and O.~Nachtmann,
  %``Optimal observables for the measurement of three gauge boson couplings
  % in e+ e- ---> W+ W-,''
  Z.\ Phys.\ C {\bf 62} (1994) 397.
  %%CITATION = ZEPYA,C62,397;%%
\bibitem{Gunion:1996vv}
  J.F.~Gunion, B.~Grzadkowski and X.-G.~He,
  %``Determining the top - anti-top and Z Z couplings of a neutral Higgs boson
  % of arbitrary CP nature at the NLC,''
  Phys.\ Rev.\ Lett.\  {\bf 77} (1996) 5172
  (hep-ph/9605326).
  %%CITATION = HEP-PH/9605326;%%  
\bibitem{Kidonakis:2013zqa}
  N.~Kidonakis,
  %``Top Quark Production,''
  arXiv:1311.0283 [hep-ph].
  %%CITATION = ARXIV:1311.0283;%%
\end{thebibliography}
\end{document}